\documentstyle[12pt]{article}

\large
\newcommand{\be}{\begin{eqnarray}}
\newcommand{\ee}{\end{eqnarray}}
\newcommand\del{\partial}

\newcommand\ie {{\it i.e. }}

\newcommand\etc{{\it etc. }}

\newcommand\half{\frac 1 2 }
\newcommand\noi {\noindent}

\begin{document}
\setlength{\baselineskip}{21pt}
\pagestyle{empty}
\vfill
\eject
\begin{flushright} USITP-93-19 \\
SUNY-NTG-93-43 \\
December 1993
\end{flushright}
\vskip 1.5cm
\centerline{\Large\bf Finite Temperature Correlators in the Schwinger Model}
\vskip 1.2 cm
\begin{center}
{\bf A. Fayyazuddin, T.H. Hansson\footnote{Supported
by the Swedish Natural Science
Research Council.} }  \\
Institute of Theoretical Physics\\
University of Stockholm \\
Box 6730, S-113 85 Stockholm, Sweden \\
\vskip 3mm
{\bf M. A. Nowak\footnote{Supported
in part by the KBN under Grant No.PB2675/2.}${}^{,3}$
, J.J.M. Verbaarschot${}^3$  and  I. Zahed\footnote{Supported
by the US Department of Energy under Grant  No. DE-FG02-88ER40388.}
 }\\
Department of Physics \\
SUNY at Stony Brook \\
Stony Brook, New York, 11794, USA \\
\end{center}
\setcounter{footnote}{0}

\vskip 1cm\noi
\centerline{\bf ABSTRACT}
\vskip .5cm
We discuss the correlation function of hadronic currents  in the Schwinger
model
at finite temperature $T$. We explicitly construct the retarded correlator
in real time and obtain analytical results for the Euclidean correlator
on a torus. Both constructions lead to
the same finite temperature spectral function. The spatial screening lengths
in the mesonic channels are related to the dynamical meson mass
$m=e/\sqrt{\pi}$ and not $2\pi T$ even in the infinite temperature limit.
The relevance of our results for the finite  temperature problem in four
dimensions is discussed.

\vfill
\eject
\pagestyle{plain}

\newcommand \etabar {\overline\eta}
\newcommand \psibar {\overline\psi}
\newcommand \thetaf [4] {\Theta\left[\begin{array}{c}#1 \\ #2
\end{array}\right] (#3 , #4) }
\newcommand \Dsl {D\!\!\!\!/}
\newcommand \dsl {{\partial}\!\!\!\!/}
\setlength{\baselineskip}{23pt}

\noindent{\bf 1. Introduction }
\renewcommand{\theequation}{1.\arabic{equation}}
\setcounter{equation}{0}
\vskip 5mm

With the possibility  to use ultrarelativistic heavy ion colliders to probe
hadronic matter at very high temperatures and/or densities it becomes
imperative to get an improved understanding of finite temperature gauge
field theories. A variety of approaches have been devised ranging
from first principle lattice simulations \cite{pet}
to approximate estimates in simplified models \cite{few}.

Just as in conventional plasma physics, a central problem in finite
temperature QCD is to find the excitation
spectrum  as a function of temperature, density,  external fields, \etc. The
spectral information in some particular channel $\alpha$ is  contained
in the spectral function $\sigma_\alpha$. Its knowledge
would, at least in principle, allow us not only to
calculate responses to external probes, but also production rates,
multiplicities, \etc.

While the spectral function at zero temperature contains poles corresponding
to bound states and resonances and cuts corresponding to particle production
thresholds, its interpretation at finite temperature is not always clear.
It is usually assumed that the spectral function at finite
temperature is dominated by hadronic quasiparticles in the low temperature
phase and various collective excitations in the high temperature phase.
However, the
nature and quantum numbers of these collective excitations are not
very well understood.

To determine the spectral function in a theory like QCD from first principles
remains an unsolved problem. Besides the purely technical/numerical problems,
the present lattice simulations of finite temperature systems
can primarily provide information about space-like correlators only.
The spectral function at finite temperature, which is related to time-like
correlators, can only be obtained if one makes very specific, and unprovable,
assumptions about its analytic structure.

Lacking suitable lattice techniques,  other methods have been used to
get spectral information. Examples are finite temperature QCD sum rule
calculations and phenomenological models (usually inspired by models for the
zero temperature vacuum), but here one is forced to
$assume$ a special shape for the spectral function (shifted poles and cuts)
and then proceed to estimate some unknown parameters like resonance
masses and  thresholds (see \cite{Shuryak-1993} for a review).
Another approach advocated by two of us
\cite{hans}, is to
use the method of dimensional reduction to calculate certain
static correlation functions at finite temperature.

The questions we would like to raise in this paper are twofold:
1. Is it meaningful to talk about spectral functions in interacting
field theories at finite temperature? 2.~To what extent can we learn about
them from finite temperature Euclidean correlators?
We will not attempt to answer these questions for
QCD, but use a model theory simple enough to obtain answers.

We consider the Schwinger model at finite temperature. In section 2 we
explicitly construct the real time correlation function for pseudoscalar
and scalar sources
(analogue of $e^+e^-$ production channel) at zero temperature and derive an
explicit relation for the zero temperature spectral function. In section 3, we
extend the analysis to finite temperature, by considering the retarded
response function, and derive an explicit expression for the finite
temperature spectral function for the Schwinger problem, using operator
techniques. In section 4, we evaluate the scalar and pseudoscalar
correlator for the Schwinger model on the torus. Exact analytical results are
obtained with the help of the properties of the Jacobi theta-functions.
The results are in full agreement with the naive bosonisation approach.
Issues related to the temporal and spatial screening masses are discussed
in section 5. In section 6, we show that the analytical continuation
of the full nonperturbative Euclidean correlators yields
temperature dependent spectral functions that are
identical to the ones derived from the real time approach.
Our conclusions are summarized in section 7. Helpful calculational
details are given in two appendices.

\vskip 1cm

\noi{\bf 2. Feynman Correlators }
\renewcommand{\theequation}{2.\arabic{equation}}
\setcounter{equation}{0}
  \vskip 5mm

The Schwinger model is quantum electrodynamics \cite{Schwinger-1962}
in one space and one time
dimension defined by the Lagrangian density
\be
{\cal L} = \frac 14 F_{\mu\nu}^2 - \bar\psi i\Dsl \psi,
\ee
where $F_{\mu\nu} =\del_\mu A_\nu - \del_\nu A_\mu$ is the Abelian field
strength and $\Dsl = \gamma_\mu(\del_\mu - ieA_\mu)$ the covariant derivative.
It is well known that at zero temperature this model is equivalent
to a theory with a single massive and free meson with  mass
$m=e/\sqrt{\pi}$ where $e$ is the electric charge of the electron
\cite{Coleman-etal-1975}. The meson
field is related to the polarization density induced by the electrons.

The mesonic Green's functions in the vacuum can be constructed using the
bosonisation techniques discussed by Coleman \cite{Coleman-1975}.
In this paper we consider the following scalar and pseudoscalar correlators,
\be
S_F(x) &=& \langle {\rm T} (\overline\psi \psi (x)
                         \overline\psi \psi (0))\rangle,
\label{scal} \nonumber\\
P_F(x) &=& \langle {\rm T} (\overline\psi i\gamma_5\psi (x)
                         \overline\psi i\gamma_5\psi (0))\rangle,
\label{pseu}
\ee
which can be readily evaluated by using the operator identity
\cite{Coleman-1975}
 \be
\overline\psi \frac 12 ( 1 \pm \gamma_5) \psi (x) =
-\left(\frac{me^C}{4\pi}\right)  : e^{\pm 2i\sqrt{\pi}\phi (x)} :,
\label{bos}
\ee
where $\phi$ is a free massive meson
and $C=0.577$ is the Euler-Mascheroni constant.
 Normal ordering with respect to the
vacuum is denoted by $:\,\, :$.
The field $\phi (x)$ can be decomposed into positive and negative
frequency components as
\be
\phi (x) = \int\frac{dp}{\sqrt{4\pi E_p}}\left(
e^{-iE_px^0 +ip x^1}a_p + e^{+iE_px^0 -ip x^1}a_p^+ \right) =
                      \phi^- (x) + \phi^+ (x),
\label{phi}
\ee
with $E_p=\sqrt{p^2 +m^2}$. The product of normal
ordered operators obtained by  inserting (\ref{bos}) in (\ref{pseu})
can be evaluated with the help of the following relation
($\epsilon_{x,0}=\pm 1$)
\be
:e^{i\epsilon_x\phi (x)} ::e^{i\epsilon_0\phi (0)}: =
e^{\epsilon_x\epsilon_0 [\phi^+ (0) , \phi^- (x) ]}
e^{i\epsilon_x\phi^+ (x) +i\epsilon_0 \phi^+ (0)}
e^{i\epsilon_x\phi^- (x) +i\epsilon_0 \phi^- (0)},
\label{camp}
\ee
which can be established  using the Baker-Campbell-Hausdorff formula.
Inserting (\ref{bos}) and (\ref{phi}) into (\ref{pseu}) and using repeatedly
(\ref{camp}) yield
\be
P_F(x) = \langle\psibar\psi\rangle^2{\rm sinh} (- 4\pi i\Delta_F (x, m^2) ),
\label{F1}
\ee
where the Feynman propagator is given by
\be
\Delta_F (x, m^2) = \theta (x^0) [\phi^- (x), \phi^+ (0)] -
         \theta (-x^0) [\phi^+ (x), \phi^- (0)],
\label{fey}
\ee
and the fermion condensate is
\be
\langle\psibar\psi\rangle = -\left(\frac{me^C}{2\pi}\right).
\ee
Similarly,
\be
S_F(x) = \langle\psibar\psi\rangle^2 \cosh (- 4\pi i\Delta_F (x, m^2) ).
\label{S1}
\ee
These results were first obtained by Casher, Kogut and
Susskind \cite{Casher-etal-}.
By expanding the hyperbolic functions in (\ref{F1}) and (\ref{S1}) we observe
that
the pseudoscalar correlator is a sum of exchanges of an odd number of massive
mesons while the scalar correlator (\ref{S1}) is given by exchange of an
even number of mesons. The scalar field carries odd parity.
In addition, the scalar correlator is  nonvanishing and equal to
$\langle\psibar\psi\rangle^2 $ in the $x\rightarrow \infty$ limit as expected
from the  cluster decomposition. The nonvanishing of
$\langle\psibar\psi\rangle^2 $ in this
model was first discussed by Loewenstein and Swieca \cite{losw}.

The spectral function at zero temperature
follows from (\ref{pseu}) by inserting a complete set of
physical states and making use of translational invariance,
\be
P_F(x) =- i\int_0^{\infty} dm^2 \sigma_P (m^2) \Delta_F (x, m^2),
\label{kall}
\ee
where we have defined ($q^0 > 0$ and $p_n^2 =m^2$)
\be
\sigma_P (q^2) = 2\pi\sum _n \delta^2 (q-p_n)
              |\langle 0|\overline\psi i\gamma_5 \psi (0) |n \rangle|^2.
\label{dens}
\ee
A comparison of (\ref{F1}) with (\ref{kall}) yields the following
spectral function for the pseudoscalar channel
\be
\sigma_P (q^2 )&=& \langle\psibar\psi\rangle^2
\left( 4\pi\delta (q^2 -m^2) \right. \nonumber \\
&+&\left. 2\pi\sum_{k=1}^{\infty} \frac{2^{2k+1}}{(2k+1)!}
\prod_{i=1}^{2k+1}\int \,
d^2p_i\delta (p_i^2-m^2 )\theta(p_i^0)
\delta^2(q-\sum_{j=1}^{2k+1} p_j )\right),\nonumber\\
\label{spec}
\ee
which was also obtained in \cite{some-russians}.
The first term corresponds to the meson pole with a strength
determined by the quark condensate.
The sum is over the $3,5,  \cdots,\infty$-particle phase space with a coupling
strength proportional to the pole strength. The spectral function
(\ref{spec}) is just the sum over the cuts of the
Feynman diagrams in the expansion of the correlator $P_F$ in powers of
$\Delta_F(x)$. A similar expression can be derived for $S_F(x)$ involving an
even exchange of pseudoscalar mesons.

\vskip 1.5cm
\noi{\bf 3. Retarded Correlators }
\renewcommand{\theequation}{3.\arabic{equation}}
\setcounter{equation}{0}
\vskip 0.5cm

The finite temperature spectral function can be obtained from the retarded
correlator. For that  consider for instance the pseudoscalar
correlator
\be
P_R (x) = \theta (x^0) \langle [ \overline\psi i\gamma_5\psi (x),
                         \overline\psi i\gamma_5\psi (0) ]\rangle_\beta,
\label{ret}
\ee
where $\langle \cdots \rangle_\beta$ denotes the Gibbs average
with respect to a heat bath of mesons of mass $m=e/\sqrt{\pi}$.
The expectation value in (\ref{ret}) can be calculated using the
operator identities (\ref{bos}) and (\ref{camp}).
In a heat bath the expectation
value $\langle \phi^+\phi^- \rangle_{\beta}$ does not vanish so that
\be
&&\langle \exp \left({i\epsilon_x\phi^+ (x) +i\epsilon_0 \phi^+ (0)}\right )
\exp \left({i\epsilon_x\phi^- (x) +i\epsilon_0 \phi^- (0)}\right)
\rangle_\beta \nonumber =\\
&&{\rm exp}\left({-\epsilon_0\epsilon_0 \langle \phi^+(0)\phi^-(0)
\rangle_\beta
     -\epsilon_x\epsilon_x \langle \phi^+(x)\phi^-(x) \rangle_\beta}
{
     -\epsilon_0\epsilon_x \langle \phi^+(0)\phi^-(x) \rangle_\beta
     -\epsilon_x\epsilon_0 \langle \phi^+(x)\phi^-(0) \rangle_\beta}\right ).
\nonumber\\
\ee
With this in mind, the retarded pseudoscalar correlator becomes
\be
P_R (x)&=&-\theta(x^0)\frac{m^2e^{2C}}{4\pi^2}
e^{-8\pi\langle \phi^+ (0)\phi^-(0) \rangle_{\beta} }\nonumber\\
&&\times \left[ {\rm sinh}\left(4\pi ([\phi^-(x),\phi^+ (0)]+ 4\pi
\langle \phi^+(x)\phi^-(0) \rangle_{\beta}+
4\pi\langle \phi^+(0)\phi^-(x) \rangle_{\beta})\right)\right. \nonumber\\
&&-\left. {\rm sinh}\left(4\pi ([\phi^-(0),\phi^+ (x)]+ 4\pi
\langle \phi^-(x)\phi^+(0) \rangle_{\beta}+
4\pi\langle \phi^+(x)\phi^-(0) \rangle_{\beta})\right)\right].\nonumber\\
\label{int2}
\ee
In the heat bath
\be
[\phi^-(x),\phi^+ (0)]
 + \langle \phi^+(x)\phi^-(0) \rangle_{\beta}+ \langle \phi^+(0)\phi^-(x)
\rangle_{\beta}
=  \int\frac{d^2p}{(2\pi)^2} \delta(p^2-m^2)(\theta(p^0)+n_p)e^{-ip\cdot
x},\nonumber\\
\label{int3}
\ee
where $n_p =1/(e^{\beta E_p} -1)$ is the
Bose-Einstein distributions for the massive bosons.
Using (\ref{int3}) in (\ref{int2}) gives
\be
P_R (x)= -\theta(x^0)
\langle \psi\psibar\rangle_\beta^2\,~& &\left [ \sinh \left(4\pi
\int\frac{d^2p}{(2\pi)^2} e^{-ip\cdot x}\delta (p^2-m^2)(\theta(p^0)+n_p)
\right)\right .  \nonumber \\
&-& \left .\sinh \left(4\pi
\int\frac{d^2p}{(2\pi)^2} e^{-ip\cdot x}\delta (p^2-m^2)(\theta(-p^0)+n_p)
\right) \right ],
\label{retf}
\ee
where
\be
\langle \psi\psibar\rangle_\beta = -\frac{m e^{C}}{2\pi}
e^{-4\pi \langle \phi^+ (0)\phi (0)^-  \rangle_{\beta}} =
-\frac{m e^{C}}{2\pi} \exp\left(-\int {dp} \frac{\,n_p }{\sqrt{p^2+m^2}}\right
)
\label{cond1}
\ee
is the temperature dependent chiral condensate, as calculated in
\cite{sachs1,manton,Smilga-1993}. A similar result holds for the retarded
scalar
correlator $S_R (x)$ with the substitution ${\rm sinh}\rightarrow {\rm cosh}$
in (\ref{retf}). Micro-causality can be exhibited explicitly
in (\ref{retf}) by rewriting the result in the form
\be
P_R (x) = \langle \psi\psibar\rangle_\beta^2\,\,{\rm sinh}(2\pi \Delta_R
(x, m^2))\,\, {\rm cosh}\left( \int \frac{d^2p}{(2\pi )^2} (1+2n_p )\,
{\rm Im} \Delta_F (p, m^2)\,\,e^{ip\cdot x}\right),
\label{cau1}
\ee
where $\Delta_R (x, m^2)$ is the retarded propagator
\be
\Delta_R (x,m^2) = \theta (x^0)\left(
[\phi^+ (x) , \phi^- (0)] + [\phi^- (x), \phi^+ (0)]\right),
\label{cau2}
\ee
and $\Delta_F (p, m^2)$ the Fourier
transform of the Feynman propagator (\ref{fey}).
Again, the retarded pseudoscalar propagator in the heat bath corresponds
to the exchange of an odd number of pseudoscalar mesons with temperature
dependent strengths. The temperature dependent strengths
follow from the bubble insertions at the vertices. Fig. 1a shows the
temperature insertions on the vertices, and Fig. 1b the diagrammatic expansion
of (\ref{cau1}). A similar result can be
derived for the scalar correlator. The answer is
\be
S_R (x) = \langle \psi\psibar\rangle_\beta^2\,\,{\rm sinh}(2\pi \Delta_R
(x, m^2))\,\, {\rm sinh}\left( \int \frac{d^2p}{(2\pi )^2} (1+2n_p )\,
{\rm Im} \Delta_F (p, m^2)\,\,e^{ip\cdot x}\right)
\label{cau3}
\ee
and involves an exchange of an even number of pseudoscalar mesons, as expected
from parity. Fig. 1c shows the corresponding diagrammatic expansion.

Using translational invariance, the retarded correlators can be related to
the pertinent finite temperature spectral function. In the pseudoscalar case
we have
\be
P_R(x) =\int d^2 q \, \frac{\theta(x^0)}{2\pi}e^{iqx}
\sigma_P(q^0, q^1)\,
\label{rets}
\ee
where the temperature dependent spectral function reads
\be
\sigma_P (q^0, q^1) = 2\pi\sum_{A,B}\delta^2(q-(q_A-q_B))(e^{-(E_A-F)/T}
-e^{-(E_B -F)/T})
\,\,|<A|\overline{\psi}i\gamma_5\psi (0) |B>|^2.\nonumber\\
\ee
The free energy of the system is denoted by $F$ and $q_A^2=m^2$.
The spectral function for the retarded correlator is obtained by
expanding the $\sinh$ in (\ref{retf}) in a power series.
To proceed the following identity is  useful
\be
\prod_{i=1}^{2k+1} (\theta(p^0_i) + n_{p_i}) -
\prod_{i=1}^{2k+1} (\theta(-p^0_i)
+n_{p_i}) = 2\sinh(\beta p^0/2) \prod_{i=1}^{2k+1} e^{\beta |p_i^0|/2} n_{p_i},
\label{ident}
\ee
where $p^0 = \sum_i p^0_i$. The proof of this identity follows immediately
if one realizes that
\be
\prod_{p^0_i> 0}(1+ n_{p_i})\prod_{p^0_i< 0} n_{p_i} -
\prod_{p^0_i> 0}n_{p_i}\prod_{p^0_i< 0}(1+ n_{p_i})=
\frac {\prod_{p^0_i> 0} e^{\beta |p^0_i|} - \prod_{p^0_i< 0} e^{\beta |p^0_i|}}
      {\prod_{p^0_i}( e^{\beta |p^0_i|} -1) }.
\ee
By comparing the expansion of the correlator and (\ref{rets})
we can read the spectral function
\be
\sigma_P (q^0, q^1 ) &=&4\pi\langle\bar \psi \psi\rangle^2_{\beta}
\sinh(\beta q_0/2)
\nonumber \\
&\times&\sum_{k=0}^{\infty} \frac{2^{2k+1}}{(2k+1)!}
\prod_{i=1}^{2k+1}\int d^2 p_i \delta(p_i^2 -m^2)\,\,e^{\beta
|p^0_i|/2} n_{p_i} \delta^2(q-\sum_1^{2k+1} p_j ).\nonumber\\
\label{specf}
\ee
A rerun of the above argument for the retarded scalar correlator gives
(following the substitution ${\rm sinh}\rightarrow {\rm cosh}$ in (\ref{retf}))
\be
\sigma_S (q^0, q^1 ) &=&4\pi\langle\bar \psi \psi\rangle^2_{\beta}
\sinh(\beta q_0/2)\nonumber \\
&\times &\sum_{k=1}^{\infty} \frac{2^{2k}}{(2k)!}
\prod_{i=1}^{2k}\int d^2 p_i \delta(p_i^2 -m^2)\,\,e^{\beta
|p^0_i|/2} n_{p_i} \delta^2(q-\sum_1^{2k} p_j ).\nonumber\\
\label{specff}
\ee
The spectral
functions at finite temperature are just the zero temperature
spectral functions with temperature dependent strengths and thermally weighted
phase space. The thresholds remain temperature independent.
This result was expected, since the bosonised version of the
model is a model with a free single massive meson.

The above results are consistent with the conventional cutting rules in real
time. Specifically, we have for the scalar ($p_2^1= q^1-p_1^1$) channel
\be
&&\sigma_S (q^0, q^1 )  = 4\pi \langle\bar\psi\psi \rangle^2_{\beta}\int
\frac{dp^1_1}{2E_12E_2}\times \nonumber\\ &&\left(
\delta (q^0-E_1-E_2) ((1+n_1)(1+n_2) -n_1n_2) +
\delta (q^0+E_1+E_2) (n_1n_2-(1+n_1)(1+n_2)) +\right .\nonumber\\&&
\left . \delta (q^0-E_1+E_2) (n_2(1+n_1)-n_1(1+n_2))+
\delta (q^0+E_1-E_2) (n_1(1+n_2)-n_2(1+n_1))\right) + \cdots,\nonumber\\
\label{w1}
\ee
where the dots refer to the processes with $4,6, \cdots$ pseudoscalar meson
exchanges.
For the pseudoscalar channel the first two terms in the
expansion of the $\sinh$-functions  in (3.5) result in
($p_3^1= q^1-p_1^1-p_2^1$)
\be
&&\sigma_P (q^0, q^1 )  = 4\pi \langle\bar\psi\psi
\rangle^2_{\beta}\delta (q^2-m^2) +
\frac{16\pi}3 \langle\bar\psi \psi \rangle^2_{\beta}\int
\frac{dp^1_1dp_2^1}{2E_12E_22E_3}\times\nonumber\\&&\left (
(\delta (q^0-E_1-E_2-E_3)-\delta (q^0+E_1+E_2+E_3))
((1+n_1)(1+n_2)(1+n_3) -n_1n_2n_3) \right .\nonumber\\&&
+\delta (q^0-E_1-E_2 +E_3) ((1+n_1)(1+n_2)n_3-n_1n_2(1+n_3))+
(3\leftrightarrow 1) + (3\leftrightarrow 2) \nonumber\\&&\left .
+\delta (q^0+E_1-E_2 +E_3) (n_1(1+n_2)n_3-(1+n_1)n_2(1+n_3))+
(2\leftrightarrow 3) + (2\leftrightarrow 1)\right) + \cdots,\nonumber\\
\label{w2}
\ee
where the dots refer to the processes with $5,7, \cdots$ pseudoscalar meson
exchanges.
Note that the pole contribution in the pseudoscalar channel remains temperature
independent, only its strength is affected by temperature. Fig. 2a shows the
various cuts contributing to (\ref{w1}) and Fig. 2b the various cuts
contributing to (\ref{w2}).
The thermal contributions to (\ref{w1}-\ref{w2}) reflect on the
exchanges of the sources with the  heat bath.
Incoming arrows are weighted by $n$, outgoing arrows are weighted by ($n+1$).
Lines on the left hand side
of the vertex denote absorption into the heat bath, lines on the right hand
side of the vertex denote emission from the heat bath.
 The lowest contribution
to the scalar channel is consistent with the result discussed by Weldon
\cite{weldon} in four dimensions for scalar $\phi^3$ theories. The
next to leading order contribution to the pseudoscalar channel is
consistent with the result discussed by Fujimoto $et$ $al$.
\cite{japan} in the context of scalar $\phi^4$ theories. The results
(\ref{specf}) and (\ref{specff}) summarize all the cuts for scalar
$\phi^n$ theories with $n=3,4,\cdots,\infty$. Their extension to four
dimensions
is immediate. The singularity structure of the spectral function at
finite temperature is much more involved than the zero temperature one.
Its structure
is physically transparent when expressed in the energy variable $q^0$
(in the heat bath frame) as opposed to the invariant variable $q^2$.

\vskip 1.5cm
\renewcommand{\theequation}{4.\arabic{equation}}
\setcounter{equation}{0}
\noi {\bf 4. Euclidean  Correlators}
\vskip 0.5cm

Most calculations in finite temperature field theory are performed using the
imaginary time or Euclidean formulation. At the perturbative level it has
been well established how these results are related to similar calculations
in real time. However, in the previous section we obtained the full {\it
nonperturbative} correlation function based on operator techniques in real time
and it is a nontrivial question whether it can be reproduced by
an Euclidean calculation. In  Euclidean space, finite
temperature corresponds to imaginary time with a periodicity $\beta = 1/T$.
The infrared divergences in the Schwinger model are regulated
by putting the space axis on a circle with circumference $L$
that at the end of the calculation is taken to infinity \cite{manton}.
Thus, we are led to study the Schwinger model on a torus.

We now outline the calculation of the scalar and pseudoscalar
correlation functions on a torus.
We will follow the methods developed in
\cite{Jay,Joos} and \cite{sachs1} and use the notation
of the latter reference
when needed. In the following,
all expectation values are defined on the torus except when stated otherwise.
The starting point is the
generating functional for (disconnected) fermionic Greens functions,
\be
Z[\eta,\etabar] = \langle \prod_{p=1}^k (\etabar|\psi_p )(\psibar_p|\eta)
     e^{-\int d^2xd^2y\, \etabar(x) S(x-y) \eta(y)} {\rm det}'(i\Dsl)
        \rangle_A  \label{part}.
\ee
The fermionic Green's function $S(x-y)$, with proper boundary conditions,
is derived in
Appendix ${\bf {\rm A}}$,
and the gauge-field average $\langle \cdots \rangle_A$ is with respect to
the usual Maxwell action $-\frac 1 2 E^2$. The functions $\psi_p$
are the zero modes,
and as explained in \cite{sachs1}, in the topological sector
corresponding to a non zero electric
flux $\Phi=2\pi k/e$,
there are precisely $|k|$ zero modes, all with chirality $\Phi/|\Phi|$. In this
sector the gauge potential can be decomposed as
\be
A_0 &=&  \frac {2\pi} \beta h_0 + \partial_1 \phi +\partial_0 \lambda -
          \frac \Phi V  x^1, \\
A_1 &=&  \frac {2\pi} L h_1 - \partial_0 \phi +\partial_1 \lambda,
\ee
where the $h_\mu$ are constant and $V=\beta L$ is the area of the torus.
The explicit expression for the $p^{th}$ zero mode ($p = 1, 2, \cdots |k|$)
with chirality $a=\pm$ can be found in \cite{sachs1}:
\be
\psi_p^a(x)  = e^{-a\phi(x)} \left(\frac {2|k|} {\beta^2 V} \right)^{\frac 1 4}
 e^{2\pi i [h_1x^1/\beta - k x^0x^1/V] }
\thetaf {x^0/L + (p-\half-h_1)/k} {kx^1/\beta + h_0} {0} {i|k|\tau}
\label{zeromode}
\ee
where $\Theta$ is the generalized theta function as defined in \cite{Mumford}.
The determinant in (\ref{part}) is over the states
orthogonal to the zero modes and is
given by \cite{sachs1}
\be
 {\rm det}(i\Dsl) = \left| \frac 1 {\eta(i\tau)} \thetaf {1+h_0}
{\half - h_1} 0 {i\tau} \right|^2
    \exp \left(\frac {m^2} 2 \int d^2x\, \phi(x) \Box \phi(x) \right)
\ee
for $k=0$ and by
\be
{\rm det}'(i\Dsl) = \left(\frac V {2|k|} \right)^{|k|/2}
 \exp \left(\frac {m^2} 2 \int d^2x\, \phi(x) \Box \phi(x) + \frac {2ek} V \int
d^2x\, \phi(x) \right)
\ee
for $k\neq 0$. Here  $\eta(i\tau)$ is Dedekind's eta function.

As in the previous section we study the connected scalar and pseudoscalar
correlation functions defined by\footnote{Our
conventions for the $\gamma$-matrices in Euclidean space are :
$[\gamma_{\mu},\gamma_{\nu}]_+ = {\bf 1}_{\mu\nu}$, $\gamma_{\mu}^+
=\gamma_{\mu}$ and $\gamma_5 =i\gamma_1\gamma_2$.}
\be
S_E(x) = \langle \psibar \psi(x) \psibar\psi(0) \rangle, \nonumber\\
P_E(x) = \langle \psibar i\gamma_5 \psi(x) \psibar i\gamma_5 \psi(0) \rangle,
\label{corfun}
\ee
where the average is according to the partition function (4.1).
It is convenient to use a chiral basis for the spinors and the fermionic
Greens functions defined by
\be
\psi_\epsilon(x) = P_\epsilon \psi(x) = \frac 1 2 (1 +
\epsilon\gamma_5)\psi(x),
\ee
where again $\epsilon=\pm 1$ and similarly for $S_{\epsilon\epsilon'}(x-y)$.
The above correlation functions can be expressed as
\be
S_E(x)   &=& 2C_{+-}(x) + 2C_{++},\nonumber\\
P_E(x)   &=& 2C_{+-}(x) - 2C_{++},
\ee
where the $C_{\epsilon\epsilon'}(x)$ are defined by
\be
C_{\epsilon\epsilon'}(x) = \langle \psibar P_\epsilon \psi(x)
\psibar P_{\epsilon '} \psi(0)\rangle.
\ee
and we have simplified (\ref{corfun}) by using the relations
\be
C_{+-}(x) = C_{-+}(x),\qquad C_{--}(x) = C_{++}(x).
\ee
{}From (\ref{part}) and the chirality of the zero modes it follows
that in fact only  $k=0$ and  $k=\pm 2$   contribute to the correlators
(\ref{corfun}).
In the trivial sector only $C_{+-}(x)$ and  $C_{-+}(x)$ are non-vanishing
whereas $C_{++}(x)$ and  $C_{--}(x)$ are non-vanishing in the sectors
$k=2$ and $k=-2$, respectively.

We start with the trivial sector where
the relevant non-vanishing correlation function reads
\be
C_{+-}(x) = \langle \psibar P_+ \psi(x) \psibar P_- \psi(0)\rangle  = -
\frac 1 Z \langle
  {\rm det}(i\Dsl)\, S_{-+}(x) S_{+-}(-x)     \rangle_A  \label{cpm1},
\ee
where $Z = Z[0,0]$ is the partition function on the torus.
To evaluate the functional integral over the gauge field
we again follow \cite{sachs1} and write
\be
\langle \cdots \rangle_A \equiv \int {\cal D}A_\mu
e^{-\half \int d^2x\, E^2(x)  }= J \sum_k
   \int_0^1 dh_0 dh_1\, \prod_{m\neq 0} d\phi_m d\lambda_m
e^{-\frac{2\pi^2 k^2}{V e^2} -\half \int d^2x \phi \Box^2\phi},
\nonumber\\
\ee
where $\phi_m$ and $\lambda_m$ are the Fourier components of the
fields $\phi$ and $\lambda$ subject to
the condition $\int d^2x\, \phi(x) = \int d^2x\, \lambda(x)=0$.
Furthermore, the Jacobian $J$ does not depend on the dynamical fields
and  cancels between
numerator and denominator in (\ref{cpm1}), and the integration over
the pure gauge $\lambda$ is trivial.
Substituting the expression (A.11) for the fermionic Greens function,
and the explicit Fourier expansion
of the generalized theta function, it is now straightforward
to perform the $h$-integrations to get
\be
C_{+-}(x)  = e^{2\pi\frac {(x^1)^2} V }\frac 1 {(2\pi\beta)^2}
\left|\frac{\vartheta_1'(0)} {\vartheta_1(z)} \right|^2 \langle e^{-2\phi(x)
+ 2\phi(0)} \rangle_S,
\ee
where $z=(x^0+ix^1)/\beta$, and  $\langle \cdots\rangle_S $
denotes the connected
Greens functions with respect to the measure
$\prod_{m\neq0} d\phi_m \exp (-S_B)$ induced by the bosonic action
\be
S_B = \half \int d^2x\, \phi(x)\Box(\Box - m^2)\phi(x).
\ee
For simplicity of notation we
have introduced the Jacobi theta function $\vartheta_1(x)
=-\theta_{11}(x, i\tau)$.
Finally using the identity \cite{Kronecker-1888}
\be
\vartheta_1'(0) = 2\pi \eta(i\tau)^3,
\ee
we obtain
\be
C_{+-}(x)  = \frac 1 {\beta^2} |\eta(i\tau)|^4 e^{2\pi\frac {(x^1)^2} V }
\left|\frac{\eta(i\tau)} {\vartheta_1(z)}
\right|^2 \langle e^{-2\phi(x) + 2\phi(0)}
\rangle_S .
\label{cpm2}
\ee
The integral over $\phi$ is Gaussian and can be expressed immediately as
\be
\langle e^{-2\phi(x) + 2\phi(0)} \rangle_S = e^{4e^2[K(x,x) - K(x,0)]},
\ee
where the two-point function $K(x,y)$ is defined by
\be
K(x,y) = \langle x| \frac 1 {\Box(\Box - m^2)}| y\rangle
= \frac 1 {m^2}[\tilde\Delta_m(x-y) -  \Delta(x-y)].
\ee
Here $\tilde\Delta_m(x)$ is the Greens function
\be
\tilde\Delta_m(x) = -\frac 1V \sum_{(k,l)\ne (0,0)}
\frac {e^{2\pi i k x^0/\beta
+2\pi i l x^1/L}}{\left( \frac {2\pi k}{\beta}\right)^2+
\left (\frac {2\pi l}{L}\right)^2 + m^2},
\ee
and $\Delta$ its
massless  counterpart. The full
massive Greens function is obtained
by adding the contribution from the constant mode,
\ie
\be
\Delta_m(x-y) = \tilde\Delta_m(x-y) - \frac 1 {Vm^2}.
\ee
Thus we have
\be
\langle e^{-2\phi(x) + 2\phi(0)} \rangle_S
    = e^{4e^2K(x,x) - 4\pi \Delta_m(x) + 4\pi \Delta(x) - \frac {4\pi} {Vm^2}}.
\label{ave}
\ee
Note that (\ref{cpm2}) apparently has a free massless singularity at
$x_\mu = 0$ since $\vartheta_1(z)$ is
odd. However, the expression (4.22) for the
massless propagator was studied by Kronecker
\cite{Kronecker-1888} with the following result
\be
\Delta(x) = \frac 1{4\pi} \log \left[
\frac{|\vartheta(z,i\tau)|^2}{\eta^2(i\tau)} e^{-\frac{2\pi (x^1)^2}{V}}
\right ],
\label{kronecker}
\ee
so this singularity cancels\footnote{The cancellation of
the singularity is expected because this model does not have massless
excitations.}.

Combining equations (\ref{cpm2}), (\ref{ave}) and (\ref{kronecker})
 and using the result
\be
\langle \psibar \psi\rangle_\beta = -\frac 2\beta e^{-\frac {2\pi} {Vm^2}}
 |\eta(i\tau)|^2 e^{2e^2K(x,x)}
\ee
from \cite{sachs1} we finally get
\be
C_{+-}(x) = \frac 1 {\beta^2} |\eta(i\tau)|^4 e^{4e^2K(x,x)
- 4\pi\Delta_m(x)} =
\frac 1 4 \langle \psibar\psi\rangle^2_\beta e^{4\pi\Delta_m(x)}.  \label{cpm3}
\ee
Note that this contribution to the scalar correlator does not saturate the
cluster decomposition according to which
\be
\lim_{x\rightarrow \infty} \langle \psibar \psi(x) \psibar\psi(0) \rangle =
   \langle \psibar\psi\rangle^2.
\ee
{}From this it is clear that the missing piece must
come from the $k= \pm 2$ sectors which we now evaluate.

According to the partition function (\ref{part}),
 the correlation function $C_{++}(x)$
reads
\be
C_{++}(x)=\frac{1}{Z} <{\rm det}'(iD\!\!\!/) |\psi_1(x)\psi_2(0)
-\psi_2(x)\psi_1(0)|^2>_A.
\ee
The average over $h_0$ and $h_1$ is performed exactly
in Appendix {\bf B} resulting in
\be
C_{++}(x) = \frac {e^{-\frac {8\pi^2}{e^2V}} }
{\beta^2} |\eta(i\tau)|^2 e^{-2\pi\frac {(x^1)^2}{V}
}  |{\vartheta_1(z)}|^2 \langle
e^{2\phi(x) + 2\phi(0)}
\rangle_S,
\ee
where we  have used that the partition function is equal to
$Z= 1/(|\eta(i\tau)|^2\sqrt{2\tau})$.
The integral over $\phi$ can be carried out in exactly the
same way as in the $k = 0$ sector. The result is
\be
\langle e^{2\phi(x) + 2\phi(0)} \rangle_S
    = e^{4e^2K(x,x) + 4\pi \Delta_m(x) - 4\pi \Delta(x) + \frac {4\pi} {Vm^2}}.
\label{ave2}
\ee
Again we use the Kronecker-identity for the massless propagator which leads
to a cancellation of the $\theta-$functions. Finally
\be
C_{++}(x) =\frac 1 {\beta^2} |\eta(i\tau)|^4 e^{4e^2K(x,x) +
4\pi\Delta_m(x)} =
\frac 1 4 \langle \psibar\psi\rangle^2_\beta e^{-4\pi\Delta_m(x)}. \label{cpp3}
\ee
Combining (\ref{cpm3}) and (\ref{cpp3}) we get
the remarkably simple  results
\be
S_E(x) = \langle \psibar\psi\rangle^2_\beta \cosh[4\pi \Delta_m(x)],
\nonumber\\
P_E(x) = \langle \psibar\psi\rangle^2_\beta \sinh[4\pi \Delta_m(x)],
\label{euc}
\ee
where $S_E$ and $P_E$ are the Euclidian counterparts of (\ref{S1}) and
(\ref{F1}), respectively.
These equations are derived for a finite circle, but as shown in \cite{sachs1}
the $L\rightarrow \infty$ limit of $\langle \psibar\psi\rangle_\beta$ is well
defined and exactly equal to the temperature dependent condensate
$\langle \psibar\psi\rangle_\beta$ appearing in section 2.
These expressions were expected from the bosonisation approach described in the
preceding sections. Indeed, (\ref{euc}) are the natural versions of the free
massive boson correlators on the torus.

\vskip 1.5cm
\noindent
{\bf 5. Screening Lengths}
\renewcommand{\theequation}{5.\arabic{equation}}
\setcounter{equation}{0}
\vskip 0.5cm

Lattice simulations in four dimensions in QCD have focused on the Euclidean
correlation functions which particular emphasis on screening lengths
\cite{detar,pet}.
It is interesting to note that the spatial screening lengths following from
(\ref{euc}) are controlled by the meson mass
$m=e/\sqrt{\pi}$ and not $2\pi T$ $whatever$ the temperature. Indeed,
for large spatial separation and at low temperatures ($T\ll m/2\pi$)
(\ref{euc}) reduces to
\be
S_E (0,x^1) - \langle \psibar\psi\rangle^2_{\beta}&\sim& 2
\langle \psibar\psi\rangle^2_{\beta} \,\,K_0^2 (m |x^1| ),\nonumber\\
P_E (0,x^1) &\sim& 2\langle \psibar\psi\rangle^2_{\beta}\,\, K_0 (m |x^1|),
\label{scree}
\ee
where $K_0$ is the McDonald function. We have used the approximation
\be
\Delta_m (0, x^1)= -\frac 12
\int\frac{dp}{2\pi} \frac{e^{ipx^1}}{\sqrt{p^2+m^2}}
\left(1+\frac 2{e^{\beta{\sqrt{p^2+m^2}}} -1}\right) \sim \frac 1{2\pi}
K_0 (m |x^1|)
\label{scree1}
\ee
valid for $T\ll m/2\pi$, and dropped terms of order $e^{-m/T}$.
In (\ref{scree}) the first result reflects on a cut
corresponding to the exchange of two massive pseudoscalars\footnote{Note
that the pre-exponential factor is essential in determining the correct
singularity structure of the spectral function.
We believe this to be the
case also in four dimensions and at low temperatures  with
$K_0 (m |x^1| )\rightarrow K_1 (m  |x_1|)$.}, and the
second on a pole. For temperatures such that $T \gg m/2\pi$,
(\ref{euc}) reduce to
\be
S_E (0,x^1) - \langle \psibar\psi\rangle^2_{\beta}&\sim& 2
\langle \psibar\psi\rangle^2_{\beta} \,\,
\left(\frac{\pi T}m\right)^2 e^{-2m |x^1|},\nonumber\\
P_E (0,x^1) &\sim& 2\langle \psibar\psi\rangle^2_{\beta}\,\,
\left(\frac{\pi T}m\right) e^{-m |x^1|},\label{screee}
\ee
where we have used $\Delta_m (0,x^1)\sim -T e^{-m |x^1|}/2m$ instead of
(\ref{scree1}). We  conclude that for any temperature, the screening length
in the pseudoscalar channel is $m$ and in the scalar channel $2m$, whatever
$T$. What is
interesting to note is that for $T \gg m/2\pi$ the finite temperature Euclidean
correlator along the spatial direction, no longer $correctly$ reproduces the
singularity structure of the real time spectral function.

If we were to
dimensionally reduce the Schwinger model, we would expect the static
correlators such as the ones described here to display screening lengths of
the order of $2\pi T$, reflecting on the free field contribution to the
correlators at high temperature. Indeed,
for a free spatial massless fermion ($\omega_n = (2n +1)\pi T$),
\be
{\cal S}(0,x^1 ) = - \frac 1{\beta}\sum_n \int \frac {dp^1}{2\pi}
e^{i p^1 x^1}\frac{\gamma^0\omega_n +\gamma^1 p^1}{\omega_n^2 +{p^1}^2}
=-\frac{i\gamma^1}{2\beta}\,\,\frac{{\rm sgn}(x^1)}{{\rm sinh} (\frac{\pi
|x^1|}{\beta})},
\label{free}
\ee
after contour integration. At large distances and/or large temperatures,
only the lowest Matsubara modes $\pm \pi T$ contribute to (\ref{free}),
resulting in
\be
{\cal S}(0,x^1 ) \sim -iT{\rm sgn}( x_1 ) \gamma^1 e^{-\pi T |x^1|},
\label{free1}
\ee
in agreement with the leading behaviour from dimensional reduction.
The free fermion contribution to the scalar and pseudoscalar Euclidean
correlators are
\be
S_E^F (0, x^1 )= && -\frac 1{2\beta^2}\,\,
\frac{1}{{\rm sinh}^2 (\frac{\pi |x^1|}{\beta})}
 \sim -2T^2 \,\, e^{-2\pi T |x^1| },\nonumber\\
P_E^F (0, x^1 )= && +\frac 1{2\beta^2}\,\,
\frac{1}{{\rm sinh}^2 (\frac{\pi |x^1|}{\beta})}
 \sim +2T^2 \,\, e^{-2\pi T |x^1| },
\label{free2}
\ee
in disagreement with (\ref{screee}). Why is that? The
reason is due to the fact that the dimensionally reduced version is plagued
with infrared divergences that makes the result (\ref{free2})
inapplicable. The model offers a counterexample to  by now $naive$ intuition.

The behaviour of (\ref{euc}) along the temporal direction at low
temperature $T\ll m/2\pi$, follows from the following form of $\Delta_m
(x^0,0)$
\be
\Delta_m (x^0,0) =-\half\int \frac{dp}{2\pi}\frac {1}
{\sqrt{p^2 +m^2}}
\frac{ e^{x_0\sqrt{p^2+m^2}} + e^{(\beta-x_0)\sqrt{p^2+m^2}} }
{e^{\beta\sqrt{p^2+m^2}} -1}
 \sim
-\frac{e^{-mx^0}}{\sqrt{2\pi m x^0}} -
\frac{e^{-m(\beta -x^0)}}{\sqrt{2\pi m (\beta -x^0)}}.\nonumber\\
\label{tlow}
\ee
The singularity structure in the scalar and pseudoscalar correlator following
from (\ref{tlow}) at very low temperature agrees with the one derived from the
spatial asymptotics, as expected from $O(2)$ invariance in the
zero temperature limit. The temperature corrections mostly affect the strengths
of the singularities. These corrections are suppressed by $e^{-m/T}$.

In the temperature range $T \gg m/2\pi$, the behavior of (\ref{euc}) along the
temporal direction is given by
\be
&&S_E (x^0,0) \sim + \langle \psibar\psi\rangle^2_{\beta} \,{\rm cosh}
\left( \frac {2\pi}{m\beta} - {\rm ln} (2-2{\rm cos} (\frac{2\pi}{\beta} x^0 ))
\right) \sim +  \frac 1{2\beta^2} \,\,\,\frac{1}{{\rm sin}^2 (\frac{\pi
x^0}{\beta} )},
\nonumber\\
&&P_E (x^0,0) \sim - \langle \psibar\psi\rangle^2_{\beta} \,{\rm sinh}
\left( \frac {2\pi}{m\beta} - {\rm ln} (2-2{\rm cos} (\frac{2\pi}{\beta} x^0 ))
\right)\sim -  \frac 1{2\beta^2} \,\,\,\frac{1}{{\rm sin}^2 (\frac{\pi
x^0}{\beta} )},
\nonumber\\
\label{teuc}
\ee
We have used the fact that at high temperature \cite{sachs1}
\be
\langle\bar\psi\psi\rangle_{\beta}\sim -2Te^{-\frac{\pi T}m}.
\label{cod2}
\ee
The exponential growth in the hyperbolic functions in (\ref{teuc}) is
balanced by the exponential fall off in the condensate.
To compare this result with the free massless fermion contribution at finite
temperature, we recall that for a free temporal massless fermion
($\omega_n = (2n +1)\pi T$)
\be
{\cal S}(x^0,0 ) = - \frac 1{\beta}\sum_n \int \frac {dp^1}{2\pi}
e^{i\omega_n x^0}
\frac{\gamma^0\omega_n +\gamma^1 p^1}{\omega_n^2 +{p^1}^2} =
-\frac {i\gamma^0}{2\beta}\,\,\frac 1{{\rm sin}(\frac{\pi x^0}{\beta} )},
\label{free4}
\ee
where the last relation follows by contour integration with $x^0\neq 0$ modulo
$\beta$. This immediately shows that at high temperature the scalar and
pseudoscalar temporal correlators (\ref{teuc}) reduce to the free fermion
behavior\footnote{ Note that
the free results (\ref{free}) and (\ref{free4}) in their
exact form are analytical continuation of each other, even though the
boundary conditions are not interchangeable for fermions.}.
By squeezing the temporal direction in
Euclidean space (high temperature) the correlator in the temporal direction
picks up exclusively the free field part. This is not the case along the
spatial direction.

\newpage
\vskip 1.5cm
\noindent
{\bf 6. Analytical Continuation}
\renewcommand{\theequation}{6.\arabic{equation}}
\setcounter{equation}{0}
\vskip 0.5cm

We now show that the Euclidean correlators (\ref{euc}) give the same spectral
functions as the ones derived from the retarded correlators. The derivation
is fully nonperturbative. The Euclidean correlators at
finite temperature are periodic. This can be readily seen by writing
(\ref{corfun}) in a Hamiltonian formalism. For instance ($x=(\tau, x^1)$)
\be
P_E(\tau, x^1) = {\rm Tr}\left( e^{-\beta (H-F)} (\theta (\tau)\,\,
\psibar i\gamma_5 \psi(x)\,\, \psibar i\gamma_5 \psi(0)  +
 \theta (-\tau)\,\,
\psibar i\gamma_5 \psi(0)\,\, \psibar i\gamma_5 \psi(x) )\right).
\label{per1}
\ee
Using $\psi (\tau +\beta ) = -e^{\beta H} \psi (\tau ) e^{-\beta H}$,
it follows
immediately that $P_E(\tau +\beta, x^1)= P_E(\tau , x^1)$. Similarly for
the scalar correlator. Thus $P_E (\tau,
x^1)$ has a discrete Fourier representation ($\omega_n = 2\pi n T$)
\be
P_E(\tau, x^1) = \frac 1{\beta} \sum_n e^{-i\omega_n\tau} P_E(\omega_n, x^1).
\label{a1}
\ee
For $0<\tau <\beta$, the Euclidean correlator (\ref{per1}) can be rewritten
in the form
\be
P_E( \tau, x^1 ) =\int dp^1 e^{ip^1x^1}
\sum_{A,B} \delta (p^1-(p_B^1-p_A^1)) e^{-\beta (E_A-F)}e^{-(E_B-E_A)\tau}
|<A|\psibar i\gamma_5\psi (0)|B>|^2,\nonumber\\
\label{per2}
\ee
following the insertion of a complete set of physical states, $p_B^2=m^2$.
Inserting (\ref{per2}) and the identity in frequency space into (\ref{a1}),
give
\be
P_E(\tau , x^1) = \frac 1{\beta}\sum_n \int \frac {d\omega dp^1}{2\pi}
e^{-i\omega_n\tau +ip^1x^1} \frac{\sigma_P (\omega, p^1)}{i\omega_n-\omega}.
\label{a2}
\ee
The sum over the Matsubara frequencies $\omega_n$ can be carried
out explicitly, using the relation (for the pinch see below)
\be
\frac 1{\beta} \sum_n e^{-i\omega_n\tau} \frac 1{i\omega_n -\omega} =
\oint_{\bf C} \frac {dz}{2\pi}\frac{e^{-iz\tau}}{iz-\omega}
\frac{e^{i\beta z}}{e^{i\beta z}-1}.
\label{per3}
\ee
The contour ${\bf C}$ encloses the real axis counterclockwise. Inserting
(\ref{per3}) into (\ref{a2}) and unwinding the contour with $0<\tau <\beta$
in mind give
\be
P_E(\tau,x^1) = -\int \frac{dp^1 d\omega}{(2\pi)} e^{ip^1x^1 - \omega\tau}
\left( 1+\frac {1}{e^{\beta\omega}-1 }\right) \sigma_P (\omega,p^1).
\label{a3}
\ee
We have deliberately disregarded the pinch singularity in (\ref{per3}) at
$\omega =0$, as it gives zero contribution in (\ref{a3})
since $\sigma_P ( 0 ,p^1) =0$.

For $L\rightarrow\infty$ the bosonic propagator $\Delta_m(x)$ can be written as
\be
\Delta_m(\tau ,x^1) = -\frac 1\beta \sum_n \int \frac{dp_1}{2\pi}
\frac{e^{ip^1 x^1 + i\omega_n \tau}}{\omega_n^2 + p_1^2 +m^2}.
\label{a4}
\ee
The sum over $n$ can be converted into an integral as in (\ref{per3}),
\be
\Delta_m (\tau , x^1) =-\oint_{\bf C} \frac{dz}{2\pi}\int \frac{dp}{2\pi}
\frac {e^{ip^1x^1}}{z^2 +p^2 +m^2}\frac{e^{iz\tau }}{e^{i\beta z}-1}.
\label{wat}
\ee
Again, the contour ${\bf C}$ encloses the real axis
counterclockwise. ${\bf C}$ can
be unwound to give contributions only from the poles $\pm i\sqrt{p^2+m^2}$
on the imaginary axis in the $z$-plane (with no pinch in this case). Thus
\be
\Delta_m(\tau ,x^1) = -\int \frac{dp_0 dp_1}{2\pi}\delta(p^2-m^2)
\frac 1{2{\rm sinh}\frac{\beta |p^0|}2}\,\,
e^{-p^0(\tau -\frac {\beta}2)  + ip^1 x^1}.
\label{a5}
\ee
The lack of manifest periodicity in $\tau$ in (\ref{a5}) is due to the fact
that while unwinding the contour ${\bf C}$ above, we have explicitly restricted
$\tau$ in the range $[0, \beta]$ to allow convergence on the circles at
infinity.
The spectral function follows from (\ref{euc}) by expanding the sinh-function
in powers of $\Delta_m$, using (\ref{a5}) and comparing with (\ref{a3}).
The result is in complete agreement with (\ref{specf}) as obtained from the
retarded correlator. This shows that the full real time spectral function can
be reconstructed from the nonperturbative Euclidean correlator at finite
temperature.

In general, the full analytical knowledge of the finite temperature Euclidean
correlator is a "dream-stuff" in interacting quantum field theories in four
dimensions. We are therefore left with approximate, and usually numerical,
estimates of the these correlators. Given this fact
of life, how do we go about reconstructing the true spectral
function? More specifically, given some knowledge of the l.h.s. of (\ref{a3}),
how do we go about the r.h.s. of (\ref{a3})?

At zero temperature, the conventional wisdom is to look at the asymptotics
of the Euclidean correlators, under the assumption that the leading singularity
in the spectral function is well separated from possible background.
At low temperature
$T\ll m/2\pi$, this rationale seems to hold for the spatial asymptotics
while most of the zero temperature structure is preserved. At temperatures
$T \gg m/2\pi$ the spatial asymptotics of the Euclidean correlator reflect
solely on screening. The tails no longer encode properly the real time
singularity structure. This is more dramatic in the temporal direction,
where the free fermion behaviour is reached at high temperature, even
in the presence of a mass gap.

Besides the asymptotics of the Euclidean correlator at finite temperature,
one can probe its zero frequency (static) or zero momentum part, as first
suggested by DeTar \cite{detar0}. Indeed, from (\ref{a2})
we see that the large $x^1$-behavior of the
static part of the Euclidean correlator,

\be
P_E (\omega_n =0, x^1 ) = -\int \frac{d\omega dp^1}{2\pi} e^{ip^1x^1}
\frac{\sigma_P (\omega , p^1 )}{\omega}
\label{per6}
\ee
reflects on the possible singularities of the spectral function
in the pseudoscalar channel and/or screening.
In the Schwinger model, the covariant dispersion
relation $\omega^2 = {p^1}^2 +m^2$ holds even at finite temperature,
tying the singularity
in $p^1$ to the singularity in $\omega$.
In general, this is not true and we would expect the singularities
(poles, resonances and thresholds) to disperse in matter.
The correlator (\ref{per6}) has been the object of interesting lattice
simulations in the context of QCD \cite{detar}.
Note that the zero momentum part
of (\ref{per6}) relates directly to the pseudoscalar susceptibility
\be
\chi_P = P_E (\omega_n =0 , p^1 =0) =-2 \int_0^{\infty} d\omega
\frac{\sigma_P (\omega ,0)}{\omega},
\label{per7}
\ee
where we have used the fact that $\sigma_P(\omega, 0)$
is and odd function of $\omega$ at zero momentum\footnote{The
zero temperature limit (\ref{dens}) should be multiplied by ${\rm
sgn}(q^0)$ to allow for antiparticles, and is consistent with this property.}.
 Susceptibilities of the type
(\ref{per7})  reflect on  the bulk properties of the theory at finite
temperature \cite{steve,prakash}.
They are important $dynamical$ parameters in the theory of
linear response. For completeness, we remark that the zero momentum part of the
Euclidean correlator at finite temperature is given by
\be
P_E (\tau , q^1=0) =\int_0^{\infty} \frac{d\omega}{2\pi}
\left( e^{\omega\tau} + e^{\omega(\beta -\tau)}\right)
\frac{\sigma_P (\omega ,0)}{1-e^{\beta\omega}} =-\int_0^{\infty}d\omega
\frac{{\rm cosh}\left(\omega(\tau-\frac{\beta}2)\right)}
{{\rm sinh}(\frac{\omega\beta}2)}\,\,\sigma_P (\omega , 0).\nonumber\\
\label{per8}
\ee
For high temperatures, the l.h.s. is dominated by the free field behaviour
which should reflect itself in $\sigma_P (\omega\sim T, 0)$.
One can think of a numerical inversion of
(\ref{per8}) (inverse Laplace transforms) as a way to probe the
time-like character of $\sigma_P (\omega, 0)$ at zero momentum. In the
Schwinger model, the validity of the inversion can be checked term by term.
This inversion is worth pursuing on the lattice for the finite temperature
Euclidean correlators in QCD.

\vskip 1cm
\noindent
{\bf 7. Conclusions}
\vskip .5cm

We have explicitly constructed the zero and finite temperature correlation
functions for the scalar and pseudoscalar channels in the context of the
Schwinger model. The finite temperature spectral functions have been obtained
directly from the retarded correlators using operator methods, and from
the Euclidean
correlators by analytical continuation. The two procedures are in agreement.

The temperature dependent spectral function in the Schwinger model reflects on
thermalized, but free mesons of mass $m=e/\sqrt{\pi}$.
The poles and thresholds are temperature
independent, with temperature dependent strengths.
The absence of interactions in the bosonized version, makes the above
results specific. In  four dimensions and for theories such as QCD, we would
expect the singularities to depend explicitly on
the temperature (pions interact in the heat bath). However, we do not have
a proof.

Our expressions for the temperature dependent spectral functions can be
written as the sum of
all possible cuttings of free scalar diagrams with an arbitrary exchange of
pseudoscalar mesons, for both the scalar and pseudoscalar channel. In their
phase space version they are directly amenable to four dimensions.
They provide compact expressions for cutting rules in real time \cite{weldon}
in the context of $\phi^n$ theories.

The exact expressions for the Euclidean correlators indicate that the screening
lengths in the Schwinger model are amenable to the meson mass $m=e/\sqrt{\pi}$
and independent of the temperature, whatever the temperature.
This is so, even though the meson is a
composite of two fermions. We strongly suspect that this result is due to
the point like character of the meson in the massless Schwinger model, and
not just the anomaly. Indeed, we conjecture that in the massive Schwinger
model, the screening lengths asymptote $2\pi T$ at high temperature. We also
conjecture that the heavy-light systems display screening lengths that
asymptote $\pi T$ in this model \cite{jac}.

We note that in the Schwinger model the screening mass and the dynamical
mass (leading singularity in the pseudoscalar channel at zero temperature)
are temperature independent. The reason is that the process
of mass generation in the model is through the $U_A(1)$ anomaly, thus protected
from matter corrections at all temperatures. This is not the case for
physical excitations in four dimensions, to the exception of the $\eta '$ mass.
This raises an interesting thought for four dimensional QCD. Could it be that
the $\eta '$ mass survives matter effects? The standard scenario based on
instanton-driven anomaly would not support this, as the instantons become
suppressed at high temperature because of Debye screening. We remark, however,
that the $\eta '$ is not point like in nature.
This point and the analogy are worth investigating,
both in instanton and non-instanton approaches to the $\eta '$ problem in
matter.

The  above considerations  suggest that possible
criteria for a weakly interacting gas
of $massless$ fermions at high temperature might be formulated in terms of
the sums,
${\cal S } = S_E (x) +P_E (x)$, or the ratios, ${\cal R}={S_E (x)}/{P_E (x)}$,
taken along the spatial or temporal direction. For free massless fermions:
${\cal S}=0$ and ${\cal R}=-1$. These
criteria hold in the presence of isospin, for a variety of combinations
with proper
isospin weighting. They might still be evaded by the Coulomb bound
state approach following from dimensional reduction as
discussed by two of us \cite{hans}. This point is worth clarifying.

The massless Schwinger model
through its severe infrared sensitivity, provides a counterexample to the
general lore on the spatial
screening lengths based on dimensional reduction in three
and four dimensions for both QED and QCD.
Although the results are specific to the  model, we think
that they are still interesting. They provide a particular realization of field
theoretical ideas at finite temperature, and show how
naive arguments could be upset by infrared singularities.

In four dimensions,
QCD is also expected to suffer from strong infrared singularities at finite
temperature. However, lattice simulations \cite{pet}
suggest that at high temperature
the screening lengths are compatible with the dimensional reduction scheme
\cite{hans,ismail}, to the exception of the pion and its chiral
partner.  To what extent these
results reflect on a free spectral function is not clear though, since
these results are still compatible with correlations in the heat bath
\cite{hans,ismail}.

Our arguments in the Schwinger model indicate that at low temperature
$T\ll m/2\pi$, the large spatial asymptotics of
the Euclidean correlators determines the leading isolated singularities
of the spectral function,
provided that the $pre$-$exponents$ are properly identified.
This identification is
rapidly lost as the temperature is increased, poles and cuts are hardly
discriminated for $T\sim m/2\pi$. If the analogy with the Schwinger model
holds, the break up temperature in the QCD context would be
$T\sim m_{\pi}/2\pi\sim 20$ MeV. A similar remark holds for the behaviour of
the Euclidean correlators along the temporal direction, except for the fact
that the latter asymptote the free fermion behaviour at high temperature
even in the presence of a mass gap.
This point suggests that the lattice calculations of
temporal correlators at high temperature \cite{boyd}
may be misleading on the
interacting aspects of the theory, since they reflect primarily on the free
field content of the correlators.

An interesting way to probe the dynamical part of the spectral function at
zero momentum is to try a numerical inversion of (\ref{per8}) for QCD. Also,
if the lattice results could be analyzed in
frequency space, then  perhaps one may think about analytical
continuations a la  Baym-Mermin \cite{baym}, assuming that one can
reconstruct singularities from a numerical output on
a discrete mesh $\omega_n =2\pi n T$. In QCD at high
temperature, it may be that the singularity is a simple cut throughout $q^0$
(analyticity aside) following the change to an ionized (plasma) state.
This simple fact is still elusive, however.
Progress in these directions may come from real time alternatives to the
present Euclidean time simulations.  It should be emphasized, however,
that the Euclidean correlators at finite temperature, are interesting
in their own right. They reflect on spacelike physics
in the form of screening and susceptibilities \cite{ismail}.

Finally, it would be interesting to see, whether finite temperature
simulations of the massless Schwinger model, reproduce the exact results
derived in this paper. Also, it would be amusing to test the ideas of
finite temperature QCD sum rules \cite{tsum}
and finite temperature instanton (caloron)
calculations \cite{tinst} in this exactly solvable model.
The extension of this work to higher correlation functions will be discussed
elsewhere \cite{jim}.

\vskip 2cm
\noindent
{\bf Note Added}
\vskip .1cm

After finishing up this work we became aware of a recent paper by A. Smilga
 "Instantons in the Schwinger Model", NSF-ITP-93-151, where the scalar
correlator is evaluated at finite temperature using instanton techniques.

\newpage

\noindent{\bf Appendix A: The Euclidean fermionic Green's function}
\renewcommand{\theequation}{A.\arabic{equation}}
\setcounter{equation}{0}
\vskip 0.5cm
In this appendix we calculate the fermionic Green's function defined by
\begin{equation}
i\Dsl\, S\left(x,y\right) = i\delta\left(x-y\right) - P\left(x,y\right),
\label{eq:green}
\end{equation}
where $\Dsl$ is the Dirac operator in the presence of a $U(1)$ gauge field
\begin{equation}
\Dsl = \gamma_{\mu}\left(\partial_{\mu} - ieA_{\mu}\right),
\end{equation}
and $P$ is the projection operator onto the space of zero modes.  We are
interested in the case when there are no zero modes ($P=0$).  Since
there are always zero modes present whenever the flux $\Phi\neq 0$ we
set $\Phi =0$
\begin{eqnarray}
A_{0} & = &\frac{2\pi h_{0}}{\beta} + \partial_{1}\phi, \nonumber \\
A_{1} & = &\frac{2\pi h_{1}}{L} -\partial_{0}\phi,
\end{eqnarray}
where $h_{0},h_{1}$ are defined modulo one.  The Dirac operator has two
zero modes (one of each chirality) at $\left(h_{0},h_{1}\right)=
\left(1/2, 0\right)$. However, these zero modes are cancelled by the
determinant
and do not give rise to singularities in the integration over $h_0$ and
$h_1$. In this section we will stay away from this point.
Notice that if we write
\begin{equation}
S\left(x,y\right) = e^{-\gamma_{5}\phi\left(x\right)}U\left(x\right)g\left(x,y
\right)U^{\dag}\left(y\right)e^{-\gamma_{5}\phi\left(y\right)},
\end{equation}
where
\begin{equation}
U\left(x\right)= \exp i2\pi\left(\frac{x^{0}h_{0}}{\beta} +
\frac{x^{1}h_{1}}{L}\right),
\end{equation}
then it follows from (~\ref{eq:green}) that
\begin{equation}
i\partial\!\!\!/g\left(x,0\right) = i\delta\left(x\right).
\label{eq:blueingreen}
\end{equation}
Because of translational invariance of the Green's function we have set $y= 0$.
The boundary conditions on the Green's functions follow from the boundary
conditions on the fundamental fields\footnote{Recall that $\Phi =0$ for the
gauge fields, so that the gauge transformation one picks up under
translation in the space direction is ${\bf 1}$.}. Specifically
\begin{eqnarray}
g\left(x^{0}+\beta, x^{1};0\right) & = &- e^{-i2\pi h_{0}}g\left(x;0\right),
\nonumber \\
g\left(x^{0}, x^{1}+L;0\right) & = & e^{-i2\pi h_{1}}g\left(x;0\right).
\label{bound}
\end{eqnarray}
{}From $\gamma_{5}S\gamma_{5} = -S$ it follows that $S_{\pm, \pm}= g_{\pm, \pm}
=0$.  Equation (~\ref{eq:blueingreen}) then reduces to
\begin{eqnarray}
+{2}{\beta}\partial_{\overline{z}}g_{-+}\left(z\right) & = &\delta^{\left(2
\right)}\left(z\right), \nonumber \\
-{2}{\beta}\partial_{{z}}g_{+-}\left(\bar{z}\right)
 & = &\delta^{\left(2
\right)}\left(\bar{z}\right),
\end{eqnarray}
where $z= \left(x^{0}+ix^{1}\right)/\beta$.
The above equations tell us that $g_{-+}$($g_{+-}$) is a  holomorphic
(anti-holomorphic) function of $z$ with a single pole at $z=0$ with residue
$1/2\pi\beta$ ($-1/2\pi\beta$).  The solution that satisfies the boundary
conditions (\ref{bound}) can be written down at once
\be
g_{-+}(z) =
\frac 1{2\pi i\beta} \sum_{m\,n} \frac{e^{2\pi i(h_0 +\frac 12)m + 2\pi i h_1
n} }{z + m + n\tau} =
 \frac 1{2\pi i\beta} {\rm Ser}(h_0+\frac 12, h_1, z, 1, \tau),
\ee
where the function Ser was introduced and studied extensively by Kronecker.
It can be expressed in terms of theta functions \cite{Kronecker-1888} resulting
in
\be
g_{-+}(z) = \frac 1{2\pi i\beta}
\frac{\vartheta_1'(0)\vartheta_1(z+h_1-i(h_0+\frac 12)\tau)}
{\vartheta_1(z)\vartheta_1(h_1-i(h_0+\frac 12)\tau)}
e^{-2\pi i z(h_0+ \frac 12)}.
\label{pole}
\ee
The function $g_{+-}(z)$ is just the complex conjugate of $g_{-+}$.
The complete Green's function follows immediately by combining this
result with the gauge transformation
\begin{eqnarray}
S_{-+}\left(x,y\right) & = &g_{-+}(z)
e^{-\phi\left(x\right)+\phi\left(0\right)}
e^{ \frac{2\pi i h_0 x^0}{\beta} +
\frac{2\pi i h_{1} x^1}{L}}.\label{gpro}
\end{eqnarray}
The function $g_{-+}(z)$ can also be constructed directly in terms of theta
functions by writing down a ratio with the boundary conditions (\ref{bound})
and the pole structure as dictated by (\ref{pole}). The generalization of
(\ref{gpro}) to a nonvanishing flux sector can be found in \cite{jim}.

\newpage

\noindent{\bf  Appendix B: Contribution from the $k=2$ sector.}
\renewcommand{\theequation}{B.\arabic{equation}}
\setcounter{equation}{0}
\vskip 0.5cm
\noindent
 From (\ref{part}), the correlation function $C_{++}(x)$
in the $k=2$ sector reads
\be
C_{++} (x)=\frac{1}{Z} <{\rm det}^,(iD\!\!\!/) |\psi_1(x)\psi_2(0)
-\psi_2(x)\psi_1(0)|^2>_A
\label{Cplusplus}
\ee
where $\psi_1$ and $\psi_2$ are the zero modes (\ref{zeromode})
for  $p=0$ and $p=1$, respectively.
To evaluate (\ref{Cplusplus}),
first, we re-express $\psi_1$ and $\psi_2$ in terms of the $\theta_{00}$ and
$\theta_{10}$ functions, respectively, using the following relations
\be
\thetaf{ a}{b} {0}{2i\tau}&=&\exp (-2\pi\tau a^2 +2\pi iab)
\theta_{00}(b+2ia\tau, 2i\tau), \nonumber \\
\thetaf{ a+\frac{1}{2}}{ b}{0}{2i\tau} &=&\exp (-2\pi\tau a^2+2\pi iab)
\theta_{10}(b+2ia\tau, 2i\tau). \nonumber \\
 \ee
We are using $(10)$ for $(\frac{1}{2}0)$,
   $(11)$ for $(\frac{1}{2}\frac{1}{2})$, etc.
Second, we apply the Caspari-Kronecker formula \cite{Krazer-1903},
\begin{equation}
\theta_{00}(2v,2i\tau) \theta_{10}(2u,2i\tau)-
\theta_{10}(2v,2i\tau) \theta_{00}(2u,2i\tau)=\theta_{11}(v+u,i\tau)
                                             \theta_{11}(v-u,i\tau)
\end{equation}
Note, that in our case the argument $(v-u)$ is $h$-independent,
so the remaining integrand $|\theta_{11}(u+v,i\tau)|^2$
 over the harmonic fields is Gaussian. Thus
\be
C_{++} (x)&=&\frac{1}{\beta^2}\frac{1}{Z} |\theta_{11}(z,i\tau)|^2
e^{-\frac{8\pi^2}{Ve^2}}<e^{2\phi(x)+2\phi(0)}>_S \nonumber \\
&\times&\int dh_0dh_1 e^{-2\pi\tau[(\frac{x_1}{L}+\frac{1}{4}-\frac{h_0}{2})^2
+(\frac{1}{4}-\frac{h_0}{2})^2]} |\theta_{11}(t,i\tau)|^2,
\ee
where $t=x_0/\beta+h_1+i\tau(x_1/L+1/2-h_0)$.
Since the integration over $h_1$ reduces the double sum in
$|\theta_{11}(t,i\tau)|^2$ to a
single one, the correlator reads
\be
C_{++} (x)&=&\frac{1}{\beta^2}\frac{1}{Z} e^{-\frac{2\pi x_1^2}{V}}
|\theta_{11}(z,i\tau)|^2
e^{-\frac{8\pi^2}{Ve^2}}<e^{2\phi(x)+2\phi(0)}>_S
\sum_n\int_0^1 dh_0 e^{-2\pi \tau (n+1 -h_0 +\frac{x^1}{L})^2} \nonumber \\
&=&\frac{1}{\sqrt{2\tau}}\frac{1}{\beta^2}\frac{1}{Z} e^{-\frac{2\pi x_1^2}{V}}
|\theta_{11}(z,i\tau)|^2
e^{-\frac{8\pi^2}{Ve^2}}<e^{2\phi(x)+2\phi(0)}>_S,
\ee
where $z=(x_0+ix_1)/\beta$ denotes the relative coordinate.

\bibliographystyle{aip}
\bibliography{vacref}

\newpage
\vskip 2cm
\noindent
{\bf  Figure Captions}
\vskip 0.5cm

\noindent{\bf Fig. 1a}. Temperature effects on the sources (a).
The dashes are temperature
insertions.

\vskip 0.3 cm
\noindent{\bf Fig. 1b}.
Diagrammatic expansion of the retarded pseudoscalar correlator
$P_E$.
The straight lines are retarded propagators $\Delta_R$, and the broken lines
the imaginary part of the Feynman propagator ${\rm Im}\Delta_F$.

\vskip 0.3 cm
\noindent{\bf Fig. 1c}.
The same as ({\bf b}) but for the retarded scalar correlator $S_E$.

\vskip 0.5cm

\noindent{\bf Fig. 2a}.
Leading processes contributing to the scalar spectral function
$\sigma_S$ in the heat bath.

\vskip 0.3cm
\noindent{\bf Fig. 2b}. Leading processes contributing
to the pseudoscalar spectral function $\sigma_P$ in the heat bath.

\end{document}